\documentclass[sigconf,table]{acmart}

\renewcommand\footnotetextcopyrightpermission[1]{} %
\acmYear{2022}
\acmPrice{15.00}

\acmConference[JCDL '22]{Proceedings of the ACM/IEEE Joint Conference on Digital Libraries in 2022}{June 20--24, 2022}{Cologne, Germany}
\acmBooktitle{JCDL '22}
\acmPrice{15.00}
\acmISBN{978-1-4503-XXXX-X/18/06} %

\usepackage{makecell}
\usepackage{multicol}
\usepackage{multirow}
\usepackage{xspace}
\usepackage{adjustbox}
\usepackage{booktabs} %
\usepackage[multiple]{footmisc}
\usepackage{xcolor}

\usepackage[pscoord]{eso-pic}%

\definecolor{PreprintA}{gray}{0.925} 
\definecolor{PreprintB}{rgb}{.63,.79,.95}

\newcommand{\preprintBanner}{
\AddToShipoutPictureFG*{\put(\LenToUnit{0.5\paperwidth},\LenToUnit{0.95\paperheight}){\makebox[0pt][c]{
\renewcommand{\arraystretch}{1.5}
\setlength{\tabcolsep}{18pt}
\rowcolors{1}{PreprintB}{PreprintA}
\begin{tabular}{|p{0.55\paperwidth}|} \hline
\textbf{Preprint from \texttt{\href{https://ostendorff.org/pub/}{https://ostendorff.org/pub/}}} \\ \hline
\footnotesize
M. Ostendorff, T. Blume, T. Ruas, B. Gipp, and G. Rehm, ``Specialized Document Embeddings for Aspect-based Similarity of Research Papers'' in \textit{Proceedings of the ACM/IEEE Joint Conference on Digital Libraries (JCDL)}, 2022. \\ \hline
\end{tabular}}}}
}

\newcommand{\extended}[2]{#1}

\newcommand{\pwc}{Papers with Code\xspace}

\newcommand{\maxaspect}[1]{\underline{#1}}

\begin{document}

\title[Specialized Document Embeddings for Aspect-based Similarity of Research Papers]{Specialized Document Embeddings for\\ Aspect-based Similarity of Research Papers}

\author{Malte Ostendorff}
\email{malte.ostendorff@dfki.de}
\affiliation{%
  \institution{DFKI GmbH}
  \city{Berlin}
  \country{Germany}
}

\author{Till Blume}
\email{till.blume@de.ey.com}
\affiliation{%
  \institution{Ernst \& Young GmbH WPG – R\&D}
  \city{Berlin}
  \country{Germany}
}

\author{Terry Ruas}
\email{ruas@uni-wuppertal.de}
\affiliation{%
    \institution{University of Wuppertal}
    \city{Wuppertal}
  \country{Germany}
}

\author{Bela Gipp}
\email{gipp@cs.uni-goettingen.de}
\affiliation{%
    \institution{University of Göttingen}
    \city{Göttingen}
  \country{Germany}
}

\author{Georg Rehm}
\email{georg.rehm@dfki.de}
\affiliation{%
  \institution{DFKI GmbH}
  \city{Berlin}
  \country{Germany}
}

\renewcommand{\shortauthors}{Ostendorff et al.}

\begin{abstract}

Document embeddings and similarity measures underpin content-based recommender systems, whereby a document is commonly represented as a single generic embedding.
However, similarity computed on single vector representations provides only one perspective on document similarity that ignores which aspects make two documents alike.
To address this limitation, aspect-based similarity measures have been developed using document segmentation or pairwise multi-class document classification.
While segmentation harms the document coherence, the pairwise classification approach scales poorly to large scale corpora.
In this paper, we treat aspect-based similarity as a classical vector similarity problem in aspect-specific embedding spaces.
We represent a document not as a single generic embedding but as multiple specialized embeddings.
Our approach avoids document segmentation and scales linearly w.r.t.\@ the corpus size.
In an empirical study, we use the \pwc corpus containing $157,606$ research papers and consider the \textit{task}, \textit{method}, and \textit{dataset} of the respective research papers as their aspects.
We compare and analyze three generic document embeddings, six specialized document embeddings and a pairwise classification baseline in the context of research paper recommendations.
As generic document embeddings, we consider FastText, SciBERT, and SPECTER.
To compute the specialized document embeddings, we compare three alternative methods inspired by retrofitting, fine-tuning, and Siamese networks. 
In our experiments, Siamese SciBERT achieved the highest scores.
Additional analyses indicate an implicit bias of the generic document embeddings towards the \textit{dataset} aspect and against the \textit{method} aspect of each research paper.
Our approach of aspect-based document embeddings mitigates potential risks arising from implicit biases by making them explicit. 
This can, for example, be used for more diverse and explainable recommendations.

\end{abstract}



\maketitle

\preprintBanner

\section{Introduction}

In content-based recommender systems and other information retrieval applications, the retrieval of semantically similar documents is often performed based on document embeddings that can be derived from the text~\citep{Le2014,Devlin2019}, citations or links~\citep{Tang2015,Han2018}, and combinations of text and citations~\cite{Cohan2020,Ostendorff2022NeighborhoodCL}. 
The similarity between documents is then calculated based on the similarity of their vector representations, e.\,g., with cosine similarity~\cite{Salton1963,Ellis1993}.
Existing approaches represent a document with a single vector in the embedding space. %
This leads to a single notion of document similarity which neglects the many meanings represented within a document, e.\,g., different arguments or sub-topics.
In the context of word embeddings,~\citet{Camacho-Collados2018} define ``the inability to discriminate among different meanings of a word'' as the meaning conflation deficiency.
While the appearance of contextualized word embeddings has solved the meaning conflation for words~\cite{Peters2018,Vaswani2017}, document embeddings still suffer from this issue.

The coarse-grained similarity assessment (similar or not) neglects the many aspects in which two documents are related. 
\citet{Goodman1972} and~\citet{Bar2011} argue the concept of similarity is an ill-defined notion unless one can say what aspects are being considered to bind the compared items.
In scientific recommender systems, the similarity is often concerned with multiple facets of the presented research, e.\,g., methods or findings~\cite{Chan2018}.
Addressing these facets individually could help tailoring recommendations for specific information needs and increasing their diversity~\cite{Kunaver2017,Ge2010}.
Especially in the scientific domain, this could help bursting filter bubbles or facilitating new discoveries \cite{Portenoy2021,Narechania2022}.  

Existing approaches derive aspect-based document similarity by splitting documents into aspect-specific segments and computing a segment-level similarity~\cite{Chan2018,Huang2020,Kobayashi2018}.
Since segmentation breaks the document coherence, our prior work \cite{Ostendorff2020c} proposes to keep documents intact and to incorporate aspect information into similarity through a pairwise document classification task.
In the prior work, we perform a pairwise multi-class classification task whereby aspects in two documents are represented with a single class label.
Pairwise document classification has been successfully demonstrated for Wikipedia articles \cite{Ostendorff2020} and research papers~\cite{Ostendorff2020c}.
However, with $\mathcal{O}(n^2)$ comparisons for a corpus of $n$ documents, the pairwise multi-class classification approach scales poorly to large scale corpora.
A quadratic complexity requires extensive computation resources, in particular in combination with other computational expensive methods, e.\,g., Transformers~\cite{Vaswani2017}.

\begin{figure}[ht]
\centering
\includegraphics[page=3,clip,width=0.95\linewidth,trim={6.5cm 2.1cm 6.5cm 2.3cm},clip]{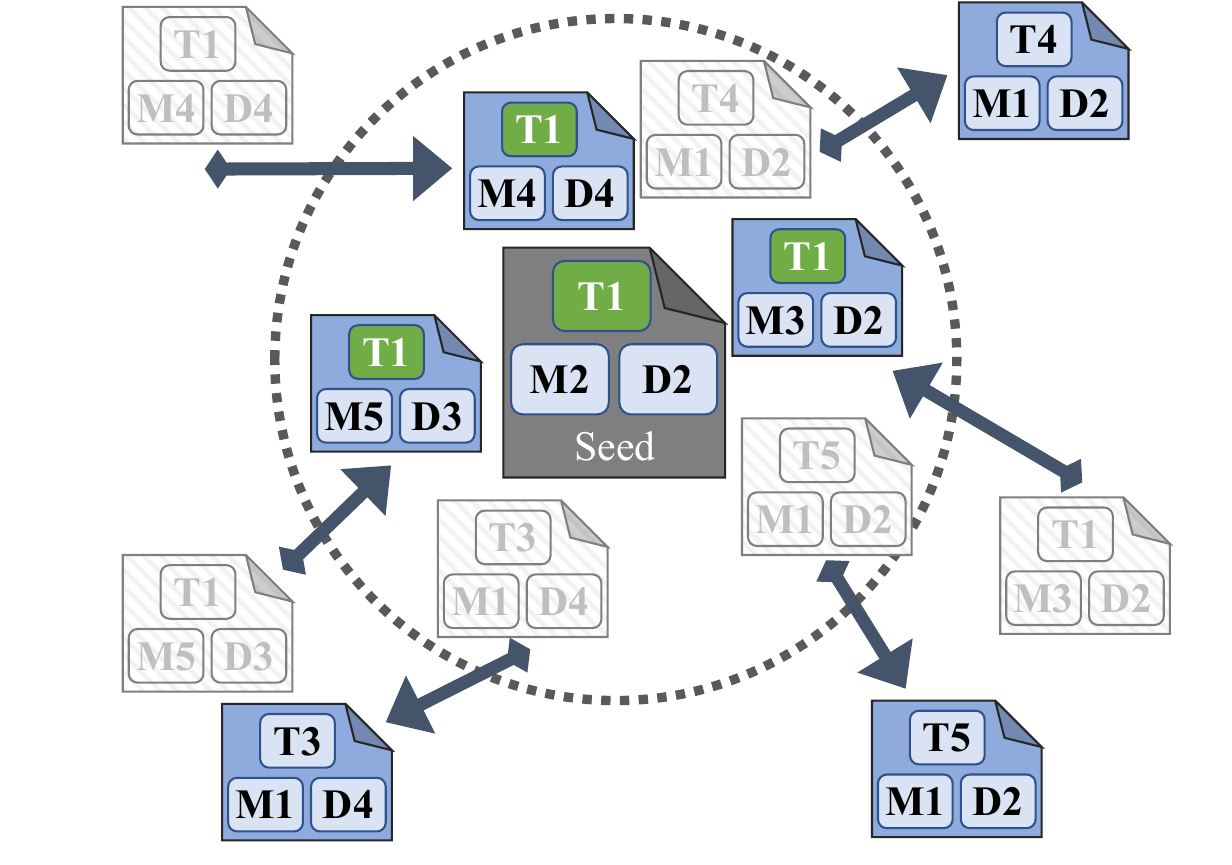}

\caption{\label{fig:idea}Papers are associated with tasks (T), methods (M), and datasets (D). With generic embeddings (gray), the $k$-nearest neighbors are papers similar in any aspect. Specializing the embeddings (blue) for the \textit{task} aspect (arrows) lets papers with the same task (T1, green) be close to each other in the embedding space.}
\end{figure}

In this paper, we present a new approach for aspect-based document similarity.
We propose to represent a document using multiple specialized embedding -- one embedding for each aspect.
We construct an aspect-specific embedding space for each aspect.
Thus, we are able to capture the similarity of documents regarding different aspects. 
We build upon the idea of specialization (sometimes referred to as retrofitting) of word embeddings~\citep{Faruqui2015,Glavas2018}.
Specialization models leverage external lexical knowledge to specialize word embedding spaces for particular constraints, e.\,g., vectors of synonyms are close to each other.
The use of multi-sense embeddings to better represent the different meaning of words is known to improve natural language understanding related tasks~\cite{LiJ15,Pilehvar2016,RuasGA19,RuasFGF20,WahleRMG21}.
We apply the idea of specialization on documents and for each aspect-specific embedding space.
Our goal is to leverage aspect information such that documents similar in a particular aspect are close to each other in the embedding space for that aspect (Figure~\ref{fig:idea}).
Thus, we refer to these embeddings as \textit{specialized} for a specific aspect in contrast to \textit{generic} embeddings that only reflect one aspect or view of a document.

Our approach keeps the documents intact as opposed to segmentation approaches~\cite{Chan2018,Huang2020,Kobayashi2018} and it addresses the scalability issues of pairwise document classification~\cite{Ostendorff2020c}.
The computational expensive encoding of aspect information is only performed once per document and aspect.
Retrieving similar documents can be done through a nearest neighbor search in each aspect-specific embedding space.
As a result, our approach has linear complexity, i.\,e., $\mathcal{O}(n)$ w.r.t.\@ to $n$ documents in the corpus.

We evaluate our approach of specializing document embeddings on a content-based recommendation task using the \pwc\footnote{\label{fn:pwc}\url{https://paperswithcode.com/}} corpus.
Research papers in \pwc are labeled with three aspects: the papers' \textit{task}, the applied \textit{method}, and the \textit{dataset} used.
We use these labels as aspects to specialize the embeddings of the research papers.
As specialization methods, we rely on existing methods but apply them in a way diverging from their original purpose. %
Namely, we evaluate retrofitting~\citep{Glavas2018} and jointly learned embeddings from Transformer fine-tuning~\cite{Beltagy2019,Cohan2020} and Siamese Transformers~\cite{Reimers2019}.
The specialized embeddings are compared against a pairwise multi-class document classification baseline and generic (non-specialized) embeddings from FastText word vectors~\citep{Bojanowski2017}, SciBERT~\citep{Beltagy2019}, and SPECTER~\cite{Cohan2020}.

In summary, our contributions are: 
    (1) We propose a new approach to aspect-based document similarity using specialized document embeddings. %
    Opposed to pairwise document classification, we treat aspect-based similarity as a classical vector similarity problem in aspect-specific embedding spaces, which improves the scalability.
    (2) We empirically evaluate three specialization methods for three aspects on a newly constructed dataset based on \pwc for the use case of research paper recommendations. 
    In our experiment, specialized embeddings improved the results in all three aspects, i.\,e., \textit{task}, \textit{method}, and \textit{dataset}.
    (3) We find that recommendations solely based on generic embeddings had an implicit bias towards the \textit{dataset} and against the \textit{method} aspect.
    (4) We demonstrate the practical use of our approach in a prototypical recommender system\footnote{\label{fn:demo}Demo \url{https://hf.co/spaces/malteos/aspect-based-paper-similarity}}.
    (5) We make our code, dataset, and models publicly available\footnote{\label{fn:github}Repository \url{https://github.com/malteos/aspect-document-embeddings}}.

\section{Related Work}

In the field of information processing, \textit{aspects} appear in various contexts and domains., e.\,g., sentiment analysis \extended{\cite{Pontiki2014}}{\cite{Hu2004,Lu2011,Pontiki2014,Do2019}}, image recommender systems \extended{\cite{Chen2017}}{\cite{Chen2017,Chen2017b}}, or reviewer matching \extended{\cite{McAuley2012}}{\cite{Karimzadehgan2008,McAuley2012}}. %
In the examples mentioned above, the goal is to associate aspect information with single items (e.\,g., products, images) or between items and users (e.\,g., review matching).
Unfortunately, very few works focus on aspect-based similarity of document pairs. %

\paragraph{Segmentation}
\citet{Chan2018} investigate aspect-based recommendations as a segmentation task.
They segment the abstracts of collaborative and social computing papers into four classes, depending on their research aspects: background, purpose, mechanism, and findings. 
Next, they represent a paper with four vectors, each derived from the corresponding segment's content. %
Computing the cosine similarity between the segment vectors allows the retrieval of similar papers for a specific aspect.
\citet{Huang2020} apply the same segmentation approach but to biomedical research papers.
\citet{Kobayashi2018} classify sections into discourse facets and build document vectors for each facet.
However, splitting documents into segments breaks the document coherence and can hurt the performance of NLP models as \citet{Gong2020} showed.
The individual segments can retain insufficient context to produce meaningful representations.
Therefore, we consider segmentation as a sub-optimal approach for aspect-based similarity.

\paragraph{Pairwise Multi-Class Document Classification}
In prior work, we propose to extend document similarity with aspect information using a pairwise multi-class document classification \cite{Ostendorff2020,Ostendorff2020c}.
The prior work evaluates the multi-class document classification approach on Wikipedia articles~\cite{Ostendorff2020} and research papers~\cite{Ostendorff2020c}.
For Wikipedia, the articles are treated as documents and Wikidata properties as labels for aspects describing their similarity~\cite{Ostendorff2020}.
For research papers, we derive aspect labels from citations and the titles of the sections in which the citations are located~\cite{Ostendorff2020c}.
Due to the inconsistent use of section titles, the titles prevent a clear distinction among aspects.
Unfortunately, no manually curated gold standard is available to date.
In both studies, variations of BERT models~\cite{Devlin2019,Beltagy2019} using a sequence pair classification setting yielded the best results~\cite{Ostendorff2020,Ostendorff2020c}.
Despite its good classification performance, pairwise classification with Transformers~\cite{Vaswani2017}, like BERT, is not suitable for large-scale similarity search applications.
Pairwise classification requires passing all possible document pairs through a Transformer model.
Thus, this approach has a quadratic complexity, as discussed also by \citet{Reimers2019}.

\paragraph{Document Embeddings}
Various methods exist to encode semantic information of documents into numerical vector representations, commonly know as embeddings.
Examples range from Bag-of-Words \cite{Harris1954}\extended{ }{over TF-IDF \cite{Jones1973}} to Paragraph Vectors~\cite{Le2014}.
Also, document embeddings from averaged word embeddings have been shown to be effective~\extended{\cite{Arora2017}}{\cite{Arora2017,RuasFGF20}}.
Recently, pretrained language models based on the Transformer architecture~\cite{Vaswani2017} have become more popular to generate embeddings based on the document text.
But also other semantic information, e.\,g., citations \extended{\citep{Han2018}}{\citep{Tang2015,Han2018}}, can be utilized for document embeddings.

\paragraph{Retrofitting}
\citet{Faruqui2015} show that word embedding learned in unsupervised fashion can be enriched with additional semantic information using retrofitting.
Retrofitting is performed in a post-processing step with external knowledge in the form of linguistic resources, such as synonyms and antonyms.
Retrofitting minimizes the distance between synonyms vectors and maximizes it between antonyms~\extended{\cite{Glavas2018}}{\cite{Mrksic2017, Glavas2018}}.
Thereby, the multi-senses of words are integrated into their vector representations.

\paragraph{Joint Learning}
Similarly, external knowledge can be directly integrated into a representation learning process.
\citet{Reimers2019} show representations from BERT~\cite{Devlin2019} can be improved with a Siamese architecture~\cite{Bromley1993} when fine-tuned on semantic textual similarity datasets. Other approaches augment pre-trained models (e.g., BART \cite{lewis2020}, RoBERTa \cite{liu2019roberta}) combining separate trained intermediate tasks and external knowledge sources to solve an additional final task, such as word sense disambiguation \cite{WahleRMG21}, paraphrase detection \cite{WahleRMG21b, WahleRFM22}, fake news detection \cite{WahleARM22}, and media bias detection \cite{spinde-etal-2021-neural-media, Spinde2022}.
Also,~\citet{Cohan2020} use citations as a pretraining objective for a scientific BERT language model.

\paragraph{Summary}

Even though the mentioned methods provide substantial contributions in document embeddings, they produce generic embeddings that represent a single view of a document's content.
This single view prevents to measure the similarity of document embeddings related to aspects. 
However, our approach aims for aspect-specialized embedding, i.\,e., for each document and for each of their aspects.
Thereby, we address issues from existing approaches for aspect-based document similarity.

\section{Methodology}

In the following, we present our approach for aspect-based document similarity and the evaluated embedding methods.

\subsection{Approach} \label{ssec:approach}

Our document embedding specialization approach, illustrated in Figure~\ref{fig:idea}, consists of two major components:
(1) aspect information for a defined set of aspects $A= \bigcup\limits_{j=1}^{n} a_j$, and (2) a specialization method that derives for any document $d_i$ in the corpus $D$ a set of $n$ specialized embeddings $\vec{d}_{i}^{(a_j)}$ for each specific aspect $a_j$ with $1 \leq j \leq n$.
The aspect information is given in the form of triples $(d_a,d_b,y^{(a_j)})$ where the label $y^{(a_j)}=\{0,1\}$ holds the binary information whether $d_a$ and $d_b$ are similar or dissimilar in aspect $a_j$.
The training objective of the specialization method is to maximize the similarity of the embeddings of those document pairs $(d_a,d_b)$ with $y^{(a_j)}=1$, i.\,e., that are similar in aspect $a_j$.

We distinguish between \textit{specialized} embeddings and \textit{generic} embeddings.
Generic embeddings can be considered aspect-free, i.\,e., $\vec{d}_{i}^{(a_1)}=\vec{d}_{i}^{(a_2)}=\vec{d}_{i}^{(a_n)}$.
\textit{Specialized} or \textit{generic} similar documents are retrieved through a $k$-nearest neighbor search using the cosine similarity of the document embeddings.
We evaluate our approach in the context of content-based recommender systems.
Therefore, we refer to the results of the nearest neighbor search as \textit{specialized} or \textit{generic} recommendations.

With this approach, we treat aspect-based similarity as a classical vector similarity problem in aspect-specific embedding spaces.
As a result, similar documents can be more efficiently retrieved as in the pairwise classification approach \cite{Ostendorff2020,Ostendorff2020c}.
Pairwise classification requires the classification of all document pairs, i.\,e., a corpus with $|D|$ documents is equivalent to $\frac{|D|*(|D|-1)}{2}$ classifications.
Thus, the pairwise classification approach has a quadratic complexity, i.\,e., $\mathcal{O}(|D|^2)$ w.r.t.\@ the number of documents $|D|$.
This quadratic complexity makes the computation infeasible even for a medium-sized corpus, in particular when Transformers are used for each classification.
Our approach computes for each document $d \in D$ and each aspect $a \in A$ one specialized document embedding $\vec{d}^{(a)}$. 
Consequently, only $|D|*|A|$ Transformer forward-passes are sufficient for inference.
Thus, our approach scales linearly w.r.t.\@ the number of documents $|D|$.
Retrieving the $k$ most similar documents can be done efficiently in the vector space using cosine similarity~\cite{Manning2008}.
For larger corpora, approximate nearest neighbor search \cite{Aumuller2017} could be also used.

\subsection{Embedding Methods}
\label{ssec:embeddings}

We evaluate the document embeddings from three base models and three specialization methods.
Besides the aspect information (Section~\ref{ssec:ground_truth}), each method utilizes the title and abstract to generate the embeddings.
We distinguish between generic and specialization methods, where the latter is divided into two categories: retrofitted and jointly learned embeddings. 
Source codes, trained models and instruction to reproduce our work are publicly available\footref{fn:github}.

\subsubsection{Generic Embeddings}
\label{sssec:generic}
We use \textit{generic} document embeddings that do not leverage any aspect information.
As base models, we rely on averaged FastText word vectors as document embeddings~\cite{Bojanowski2017}, document embeddings from SciBERT~\cite{Beltagy2019}\footnote{For SciBERT, we apply mean-pooling, i.\,e., a document vector is the mean of the hidden-states of the last layer of the SciBERT model. Documents embeddings from the \texttt{[CLS]}-token yielded significantly lower results, e.\,g., 0.001 MAP for the \textit{task} aspect).}, and SPECTER~\cite{Cohan2020}.
SPECTER and SciBERT are BERT-inspired models~\cite{Devlin2019} pretrained on scientific literature. 
In contrast to SciBERT, SPECTER uses citation prediction as an additional pretraining objective.
SciBERT and SPECTER are used as published by their authors without any fine-tuning on our corpus and in their \texttt{BASE}-version. %

\subsubsection{Retrofitted Embeddings} \label{sssec:retrofitted}
Retrofitting refers to the postprocessing of existing embeddings such that they fit predefined constraints~\cite{Faruqui2015}.
Constrains, e.\,g., synonyms or antonyms, define which vectors should be close or apart.
For our experiments, we retrofit all generic embeddings with Explicit Retrofitting (ER) as proposed by~\citet{Glavas2018}.
In contrast to other retrofitting methods~\cite{Faruqui2015}, ER generalizes to unseen vectors for which no predefined constraints exist. 
An ER model can be learned on a subset for which constraints exist (training set) and, then, be applied on all remaining embeddings (test set).
The training constraints are the positive samples in the same fashion as the synonyms are used for the retrofitting of words.

\subsubsection{Jointly Learned Embeddings}  \label{sssec:joint}
We refer to this category as jointly learned embeddings since aspect information is integrated into the representation learning process.
Aspect-based embeddings are directly generated from textual input (title and abstract of a paper).
We fine-tune SPECTER and SciBERT in a sequence-pair setup on positive and negative samples from our training set.
The input is a pair of two papers separated with a \texttt{[SEP]}-token.
The sequence pair is subject to a binary classification (similar in aspect or not).
To derive embeddings for the test set, we use only a single paper as input to SPECTER and SciBERT.
Aside from SPECTER and SciBERT, we also test a Siamese network based on SciBERT (see Sentence-BERT~\cite{Reimers2019}).
Siamese-SciBERT uses a Siamese architecture~\cite{Bromley1993}, in which the paper pair is separately fed as an input, their representations are concatenated, and then classified.\footnote{For Siamese-SciBERT, we experimented with different loss functions and found the Multiple Negative Ranking Loss~\cite{Henderson2017}, with only positive samples from the train set, yielded the best results for our data.}

\section{Experiments}

For our experiments, we use the three generic embeddings Avg.~FastText, SciBERT, SPECTER (see Section~\ref{sssec:generic}).
As specialization methods, we retrofit the three generic embeddings, and also jointly learn specialized embeddings with Transformer fine-tuning and Siamese Transformers (see Section~\ref{sssec:retrofitted} and \ref{sssec:joint}).
Furthermore, we use the pairwise classification approach as a baseline.

\subsection{Corpus}

Our approach requires information about aspects that make a document pair similar.
To the best of our knowledge, no appropriate dataset for the problem of aspect-based similarity is publicly available as they lack either quantity or quality.
\citet{Chan2018} provide a dataset that is too small in size for a machine learning approach.
In our prior work \cite{Ostendorff2020c}, we rely on citations and section titles as a training signal.
However, section titles are inconsistently used and, therefore, prevent a clear distinction among aspects.

\pwc hosts a hand-curated collection of research papers in the machine learning domain~\cite{Kardas2020}. 
In addition to metadata on authors or bibliography, each research paper is labeled with the \textit{task} a paper is focusing on, the papers' \textit{method}, and the \textit{dataset} used. 
We use these labels as aspects, $A=\{\textrm{\textit{task}},\textrm{\textit{method}},\textrm{\textit{dataset}}\}$,  as they address different information needs that are beneficial for research paper recommender systems~\cite{Chan2018}.
For example, \textit{\citet{Beltagy2019}} and \textit{\citet{Cohan2020}} are labeled with \textit{BERT} \textit{\cite{Devlin2019}} as their \textit{method}. 
Thus, we consider the pair of \textit{\citet{Beltagy2019}} and \textit{\citet{Cohan2020}} as similar regarding the \textit{method} aspect.
Other aspect labels are for example:

\begin{itemize}
    \item \textbf{Tasks:}  Low-Rank Matrix Completion, Q-Learning, Quantization, Speaker Recognition, Object Detection
    \item \textbf{Methods:} Residual Connection, Tanh Activation, Multi-Head Attention, LSTM, Transformer
    \item \textbf{Datasets:} Atari 2600 Atlantis, Cityscapes, SOP, MS MARCO, Labeled Faces in the Wild
\end{itemize}

\subsection{Ground truth}
\label{ssec:ground_truth}

The used \pwc corpus contains in total 157,606 unique papers%
.
For each aspect, we construct separated ground truths containing positive and negative samples.
Positive samples are unique unordered paper pairs with the same label, i.\,e., $y=1$. %
For each label, the number of pairs is ${L \choose 2}$ where $L$ is the number of papers per label.
Negative samples are randomly sampled paper pairs without the same label, i.\,e., $y=0$.
The number of negative samples is $50\%$ of the number of positive samples. 
Some labels are too frequent in the corpus, e.\,g., the \textit{method} label \textit{Softmax} is assigned to 5,324 papers. 
To ensure the specificity of aspect information, we discard all labels which are assigned to more than 100 papers. 
The removal of too frequent labels increases the task's difficulty and ensures an appropriate dataset size. 
The dataset would become too large otherwise, e.\,g., \textit{Softmax} alone would account for 1.2M paper pairs.
We conduct our experiments as 4-fold cross-validation and split the data into 75\% training and 25\% test papers.
The resulting ground truth consists of on average 
of 1,227,058 \textit{task}, 284,193 \textit{method}, and 58,984 \textit{dataset} paper pairs.

\begin{table}[h]
\renewcommand{\arraystretch}{0.9} %
\centering
\small

\caption{Ground truth for each aspect}\label{tab:dataset}
\begin{tabular}{lrrrr}
\toprule
\textbf{Aspect} 
&  \textbf{Papers} 
&  \textbf{Labels} 
&  \textbf{Avg.~papers per label} \\
\midrule
Task    
&  154,350 
&    1,421 
&              17.9 \\
Method  
&  108,687 
&     788 
&              12.4 \\
Dataset 
&   37,604 
&    1,743 
&               5.6 \\
\bottomrule
\end{tabular}

\end{table}

\subsection{Baseline} \label{ssec:baseline}

To compare our approach with prior work, we use the pairwise multi-class classification approach as a baseline \cite{Ostendorff2020c}.
We train a pairwise classification model based on SPECTER.
We selected SPECTER over SciBERT as its generic version outperformed SciBERT.
With a document pair as input, the model predicts the probability distribution over the aspect labels. 
The pairwise approach is not directly applicable on our dataset as its quadratic complexity would require the classification of 1.3 billion document pairs.
To reduce the number of candidate pairs, we first retrieve the $n=300$ nearest neighbors $d_n$ for any seed document $d_s$ based on the generic SPECTER embeddings.
The pairs of seed and neighbor documents $(d_s,d_n)$ are selected as candidates for the classifier.
This candidate filtering reduces the number of classifications to 11.3 million document pairs.

\subsection{Evaluation Methodology}

Each of the $n$ aspects is evaluated separately ($n$ train, $n$ test sets). 
All documents from the test set are used as seeds.
For a given aspect $a_j$ and the vector $\vec{d}_s^{(a_j)}$ of seed $d_s$, we retrieve $k$ candidate documents, with a $k$ nearest neighbor search~\cite{Cover1967}. %
The similarity of documents is computed as the cosine similarity of their vectors~\cite{Salton1963}.
The only exception is the pairwise baseline approach, for which the predicted class probabilities are used instead of cosine similarity.
A candidate document $d_c$ is relevant for the seed $d_s$ if they are associated with the same label for aspect $a_j$, i.\,e., $(d_s,d_c,y^{(a_j)}=1)$ is part of the ground truth.
We compute precision, recall, mean average precision, and mean reciprocal rank based on this relevance definition \cite{Manning2008}.

\section{Experimental Evaluation}

In the following, we present our experimental results.
We start with the evaluation of the pairwise approach baseline and continue with the comparison of all aspect-similarity methods, analyze the differences between generic and specialized embeddings, and finally verify our findings with qualitative examples.

\subsection{Pairwise Baseline Evaluation}

In order to retrieve similar documents with the pairwise approach, we first need to train a classification model that can be separately evaluated on the test set.
Table~\ref{tab:pairwise} shows the classification performance of Pairwise SPECTER in terms of precision, recall, and F1-score.
With a micro F1-score of $0.74$, the performance is comparable the previous experiments \cite{Ostendorff2020c}.
A discrepancy can be seen between the aspects.
For \textit{task} the F1-scores are the highest with $0.84$, followed by \textit{method} with $0.50$.
The worst performance yields the \textit{dataset} aspect with an F1-score of only $0.16$.

\begin{table}[!h]
\renewcommand{\arraystretch}{0.9} %
\centering
\small

\caption{\label{tab:pairwise} Classification report for Pairwise SPECTER.}
\begin{tabular}{lrrrr}
\toprule
\textbf{Aspect $\downarrow$ / Metric $\rightarrow$ } 
&  \textbf{Precision} &  \textbf{Recall} &  \textbf{F1-Score} 
\\
\midrule
Task   & 0.88  & 0.81& 0.84\\
Method& 0.56& 0.46& 0.50\\
Dataset& 0.11& 0.33& 0.16\\
\midrule 
Micro Avg.& 0.79  &    0.74     & 0.76 \\
Macro Avg.& 0.52  &    0.35 &     0.50\\
\bottomrule
\end{tabular}
\end{table}

\begin{figure}[ht]
\centering
\includegraphics[page=1,clip,width=0.95\linewidth,trim={0.cm 0.cm 0.cm 0.cm},clip]{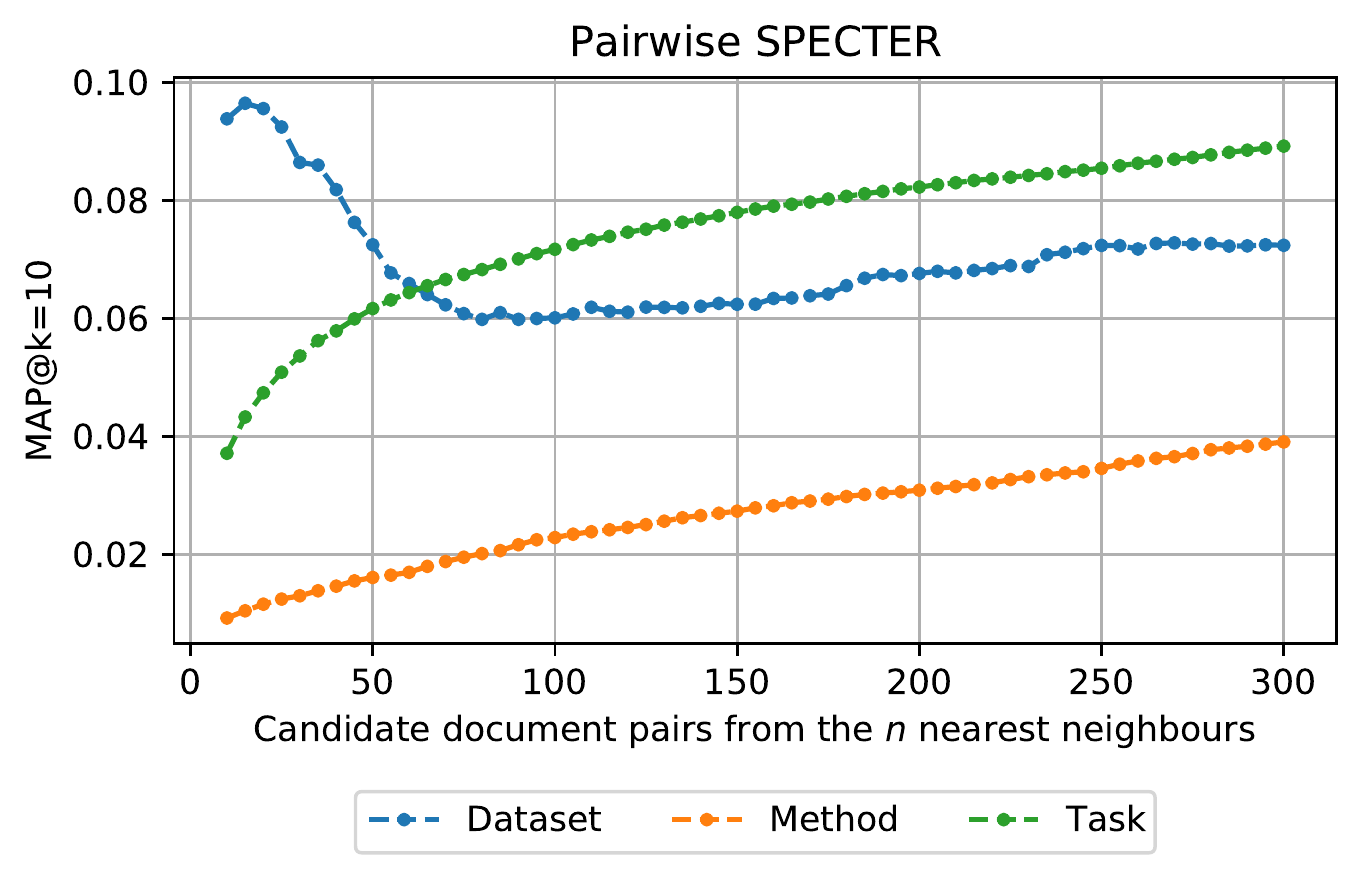}

\caption{\label{fig:pairwise}Performance of Pairwise SPECTER in terms MAP@k=10 depending on the candidate filtering for different $n$ nearest neighbors.}
\end{figure}

To make the pairwise approach applicable to our dataset, we introduced an artificial constraint since the prediction for all document pairs is not possible due to the quadratic complexity and limited resources.
We retrieve the $n=300$ nearest neighbors based on generic SPECTER to filter for candidate pairs for that we predict the aspect labels.
As this constraint potentially harms the performance, we plot Pairwise SPECTER's performance as MAP@k=10 depending on the size of nearest neighbor filter in Figure~\ref{fig:pairwise}.
The performance generally increases as $n$ increases.
However, the larger $n$ the smaller the increase is.
Thus, we expect the performance not to increase significantly for large $n$. %
The high MAP for the \textit{dataset} aspect and small $n$ is due to the good performance of generic SPECTER for this aspect.

\subsection{Aspect-based Similarity Evaluation}

\begin{figure*}[ht]
\centering
\includegraphics[page=1,clip,width=0.95\linewidth,trim={0cm 2.85cm 0cm 0cm},clip]{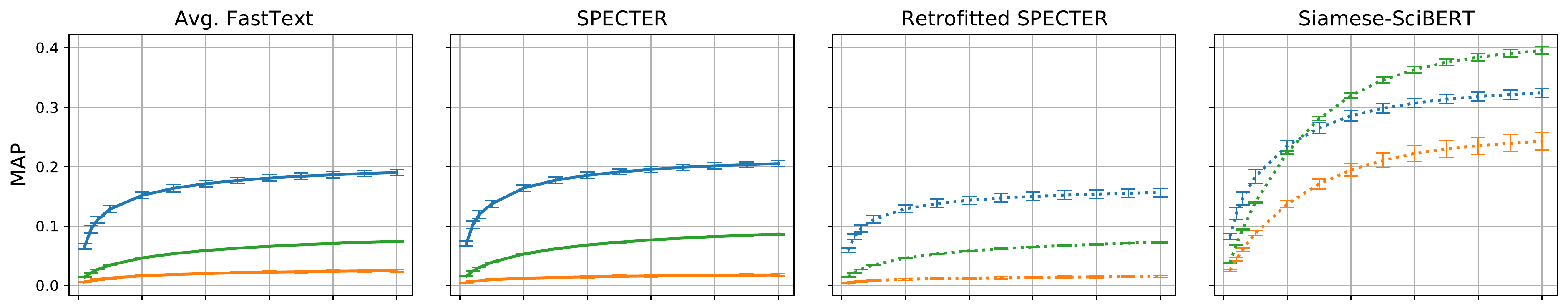}
\includegraphics[page=1,clip,width=0.95\linewidth,trim={0cm 0cm 0cm 0.7cm},clip]{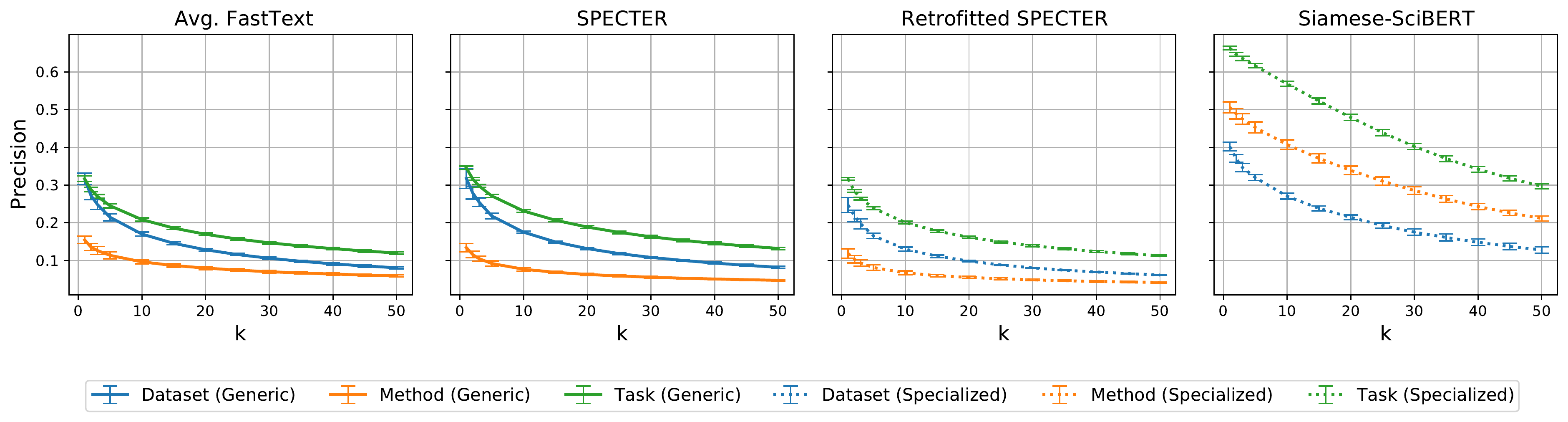}

\caption{\label{fig:map_per_k}
Precision and MAP@k for two generic (Avg.~FastText and SPECTER) and two specialized embeddings (Retrofitted SPECTER and Siamese-SciBERT). 
For generic embeddings, each line presents the scores of the generic method evaluated on different aspect-datasets. 
For specialized embeddings, a line presents a separately trained model. 
Generic embeddings and retrofitted SPECTER yield similar results on different $k$ and aspects, while for Siamese-SciBERT, the \textit{task} aspect yields a higher MAP compared to \textit{dataset} for $k>15$. 
}
\end{figure*}

Table~\ref{tab:overall_results} presents the overall results based on the most $k=10$ similar documents from each method.
Results for other $k$ values are depicted in Figure~\ref{fig:map_per_k}.
In the following, unless stated otherwise, we refer to the MAP results since it takes the rank of multiple relevant candidates into account.

Siamese-SciBERT is for all metrics and aspects the best method by a large margin.
Among the generic embeddings, SPECTER is on average better than Avg.~FastText. 
For \textit{task} and \textit{dataset}, SPECTER outperforms Avg.~FastText, while for \textit{method} the opposite is the case.
SciBERT yields the lowest scores in the generic category. 
As~\citet{Reimers2019} showed, BERT-based embeddings perform poorly without task-specific fine-tuning.
Even the computational less complex Avg.~FastText outperforms SciBERT.
Despite requiring the largest computational effort, the Pairwise SPECTER baseline yields only the second-best scores for \textit{task} and \textit{method} while for \textit{datasets} the scores are even the fourth-lowest.

\begin{table*}[ht]

\renewcommand{\arraystretch}{1.2} %

\caption{Overall results for the most $k=10$ similar documents for nine embedding methods and the Pairwise SPECTER baseline. 
Precision (P), recall (R), mean reciprocal rank (MRR), mean average precision (MAP) are reported as average over a 4-cross-validation. 
The highest score among aspects in each metric is underlined for the individual method, and \textbf{bold} shows the highest score among methods for a single metric.
Fine-tuned Siamese-SciBERT yields the best results. %
\label{tab:overall_results}}

\begin{tabular}{cl cccc cccc cccc}

\toprule
\multicolumn{2}{l}{\textbf{Aspects $\rightarrow$}} & \multicolumn{4}{c}{\textit{Task}} & \multicolumn{4}{c}{\textit{Method}} & \multicolumn{4}{c}{\textit{Dataset}} \\

\cmidrule(lr){3-6}
\cmidrule(lr){7-10}
\cmidrule(lr){11-14}

\multicolumn{2}{l}{\textbf{Methods $\downarrow$}} &      \textbf{P} &      \textbf{R} &    \textbf{MRR} &    \textbf{MAP} &      \textbf{P} &      \textbf{R} &    \textbf{MRR} &    \textbf{MAP} &       \textbf{P} &      \textbf{R} &    \textbf{MRR} &    \textbf{MAP} \\

\midrule

\multicolumn{2}{l}{Pairwise SPECTER baseline \cite{Ostendorff2020c}}    
&  \maxaspect{0.298} &  0.110 &  \maxaspect{0.545} &  \maxaspect{0.089} &  0.152 &  0.048 &  0.400 &  0.039 &   0.124 &  \maxaspect{0.119} &  0.316 &  0.072 \\

\midrule

\multirow{3}{*}{\rotatebox[origin=c]{90}{Generic}} & Avg.~FastText             
&  \maxaspect{0.208} &  0.071 &  0.419 &  0.046 
&  0.096 &  0.029 &  0.233 &  0.016 
&   0.170 &  \maxaspect{0.260} &  \maxaspect{0.439} &  \maxaspect{0.152} \\

& SPECTER                   
&  \maxaspect{0.231} &  0.080 &  \maxaspect{0.448} &  0.053 
&  0.077 &  0.023 &  0.205 &  0.012 
&   0.175 &  \maxaspect{0.277} &  0.446 &  \maxaspect{0.164} \\

& SciBERT                   
&  \maxaspect{0.083} &  0.027 &  0.241 &  0.015 
&  0.044 &  0.012 &  0.142 &  0.006 
&   0.079 &  \maxaspect{0.112} &  \maxaspect{0.251} &  \maxaspect{0.059}  \\

\cmidrule(lr){1-14}

\multirow{6}{*}{\rotatebox[origin=c]{90}{Specialized}} 
& Retrofitted Avg.~FastText 
&  \maxaspect{0.233} &  0.081 &  0.445 &  0.054 
&  0.133 &  0.040 &  0.294 &  0.024 
&   0.202 &  \maxaspect{0.290} &  \maxaspect{0.481} &  \maxaspect{0.174} \\

 & Retrofitted SPECTER       
 &  \maxaspect{0.201} &  0.071 &  \maxaspect{0.414} &  0.046 
 &  0.067 &  0.020 &  0.186 &  0.010 
 &   0.130 &  \maxaspect{0.227} &  0.364 &  \maxaspect{0.129} \\

 & Retrofitted SciBERT       
 &  \maxaspect{0.106} &  0.035 &  0.284 &  0.019 
 &  0.067 &  0.018 &  0.189 &  0.009 
 &   0.103 &  \maxaspect{0.140} &  \maxaspect{0.304} &  \maxaspect{0.073} \\

\cmidrule(lr){2-14}

 & Fine-tuned SPECTER        
 &  \maxaspect{0.279} &  0.095 &  \maxaspect{0.497} &  0.067 
 &  0.063 &  0.017 &  0.171 &  0.010 
 &   0.092 &  \maxaspect{0.134} &  0.279 &  \maxaspect{0.070} \\

 & Fine-tuned SciBERT        
 &  \maxaspect{0.091} &  0.031 &  \maxaspect{0.258} &  0.020 
 &  0.052 &  0.013 &  0.156 &  0.007 
 &   0.070 &  \maxaspect{0.088} &  0.224 &  \maxaspect{0.045} \\

 & Fine-tuned Siamese-SciBERT         
 &  \textbf{\maxaspect{0.569}} &  \textbf{0.242} &  \textbf{\maxaspect{0.708}} &  \textbf{0.224} 
 &  \textbf{0.407} &  \textbf{0.168} &  \textbf{0.588} &  \textbf{0.137} 
 &  \textbf{0.270} &  \textbf{\maxaspect{0.374}} &  \textbf{0.533} &  \textbf{\maxaspect{0.235}}  \\
\bottomrule
\end{tabular}

\end{table*}

The retrofitting approach~\cite{Glavas2018} has a mixed effect on the performance.
For Avg.~FastText and SciBERT, the retrofitting increases all scores (on average +26\% MAP for Avg.~FastText, +34\% MAP for SciBERT), while for SPECTER the retrofitting decreases the performance compared to its generic version (on average -16\% MAP).
The fine-tuning of SPECTER and SciBERT has a different effect depending on the aspects.
Compared to its generic counterpart, fine-tuned SPECTER's MAP score is 25\% higher for the \textit{task} aspect but 57\% lower for the \textit{dataset} aspect.
For SciBERT, the fine-tuning also decreases its MAP score by 23\% for the \textit{dataset} aspect.
Moreover, we do not only see performance differences between the methods but also between the aspects.
All methods yield the highest precision for \textit{task}, whereas recall and MAP are the highest for \textit{dataset}. %
A high MRR can be found for \textit{task} and \textit{dataset}, while the \textit{method} aspect shows the lowest scores throughout all metrics.
The poor \textit{method} results can be partially attributed to the unbalanced distribution of the aspects (Section~\ref{ssec:ground_truth}).
Most samples are available for \textit{task}, explaining its good performance compared to \textit{method}.
However, \textit{dataset} has the least number of samples but still outperforms \textit{method}.
As we specialize the embeddings, we also notice a decrease in performance difference between the aspects.
While SPECTER has a high MAP difference from \textit{dataset} to \textit{method} (92\%) and from \textit{dataset} to \textit{task} (68\%), the same difference is lower for Siamese-SciBERT (42\% and 5\% respectively).
The better the specialization effect the lower is the performance gap the between aspects.

To analyze the aspect-specific performance, Figure~\ref{fig:map_per_k} depicts the performance ranking as MAP and precision for different $k$ values for Avg.~FastText, SPECTER, Retrofitted SPECTER, and Siamese-SciBERT.
The performance among the aspect remains stable independent of $k$ for all methods, except Siamese-SciBERT.
With Siamese-SciBERT, the \textit{task} aspect yields a higher MAP than \textit{dataset} for $k>15$.
In terms of precision, Siamese-SciBERT is another exception since the precision of \textit{method} is higher than in \textit{dataset}.
For all other methds, \textit{method} has the lowest precision.

In summary, Siamese-SciBERT achieves, for all metrics and aspects, the highest scores.
Thus, we consider Siamese-SciBERT the best method out of the analyzed methods to handle specialized embeddings even outperforming the Pairwise SPECTER baseline. %

\subsection{Specialization Evaluation}

The performance discrepancy among the aspects could indicate a systematic difference between the documents retrieved through the similarity of generic embeddings and the specialized ones.
Therefore, we conduct an additional experiment on their overlap. 
We use the trained models from Table~\ref{tab:overall_results} but infer vectors for all documents in the whole corpus.
Then, retrieve $k=50$ recommendations and count the overlap between each method's nearest neighbors on a seed-level.
The large $k$ value is selected to increase the chance of overlapping retrieved documents.
Table~\ref{tab:overlap} presents the intersection ratio between the generic retrieved documents from Avg.~FastText and SPECTER, and the specialized ones from Siamese-SciBERT.
For the remaining methods, we report the intersection in the supplemental materials\footref{fn:github}.
The lower the overlap, the more distinct the recommendations are from each other.

On the one hand, most overlaps can be found between Avg.~FastText and SPECTER.
This suggests little difference within the generic retrieved documents.
On the other hand, Siamese-SciBERT's \textit{method}-specific recommendations overlap the least with the generic ones.
The discrepancy among the aspects is significant. 
Compared to SPECTER, Siamese-SciBERT has an overlap of 12\%, 5\%, and 17\% for \textit{task}, \textit{method}, and \textit{dataset} respectively.
Thus, indicating \textit{dataset}-specific recommendations are overrepresented in generic recommendations, while \textit{method}-specific ones are underrepresented.

\begin{table}[!h]

\centering
\renewcommand{\arraystretch}{0.9} %

\caption{Intersection of $k=50$ recommendations from A and B. 
Most overlap between generic methods (Avg.~FastText and SPECTER).
Only 5\% of Siamese-SciBERT's \textit{method} recommendations also also retrieved by generic methods. \label{tab:overlap}}

\begin{tabular}{llr}
\toprule
\textbf{Recommendations A }       &     \textbf{Recommendations  B }          &  \textbf{A$\cap$B} \\
\midrule
Avg.~FastText & SPECTER &                 0.29 \\
        & Siamese-SciBERT\textsuperscript{(Task)} &                 0.11 \\
        & Siamese-SciBERT\textsuperscript{(Method)} &                 0.05 \\
        & Siamese-SciBERT\textsuperscript{(Dataset)} &                 0.14 \\
        \cmidrule(lr){1-3}
        
SPECTER 
& Siamese-SciBert\textsuperscript{(Task)} &                 0.12 \\
& Siamese-SciBert\textsuperscript{(Method)} &                 0.05 \\
& Siamese-SciBert\textsuperscript{(Dataset)} &                 0.17 \\

\bottomrule
\end{tabular}

\end{table}

\section{Qualitative Verification}

Considering the quantitative findings, we also qualitatively analyze randomly sampled seed papers and their most similar documents in the context of research paper recommendations.
Table~\ref{tab:examples} presents one of these samples with its top-$k=3$ recommendations.
Generic recommendations are taken from SPECTER and \textit{task}-, \textit{method}-, and \textit{dataset}-specific ones from Siamese-SciBERT.
For other examples, we provide a Web-based demo to browse the recommendations for all papers from the dataset\footref{fn:demo}.

\begin{table*}
\centering

\caption{Example recommendations from SPECTER (generic) and Siamese-SciBERT (aspect-specific) for the seed \textit{``Data augmentation for low resource sentiment analysis using generative adversarial networks''} by \citet{Gupta2019} \label{tab:examples}}

\resizebox{0.99\textwidth}{!}{
\renewcommand{\arraystretch}{1.5} %
\begin{tabular}{cp{0.22\linewidth}p{0.22\linewidth}p{0.22\linewidth}p{0.22\linewidth}}

& \textbf{Generic} & \textbf{Task} & \textbf{Method} & \textbf{Dataset} \\ \hline

1 
& Adversarial Training for Aspect-Based Senti. Analysis with BERT \cite{Karimi2020} 
& Not Enough Data? Deep Learning to the Rescue!~\cite{Anaby-Tavor2019}
& DNA Methylation Data to Predict Suicidal and Non-Suicidal Deaths: A ML. Approach~\cite{Zahan2020}
& Semi-Supervised and Transfer Learning Approaches for Low Resource Senti. Class.~\cite{Gupta2018} \\

\hline

2
&  Emotion Classification with Data Augmentation Using Generative Adversarial Networks~\cite{Zhu2018}
& Towards better detection of spear-phishing emails ~\cite{Regina2020}
& Company Class. using Machine Learning~\cite{Husmann2020}
& Affection Driven Neural Networks for Senti. Analysis~\cite{Xiang2020} \\

\hline

3 
& Hierarchical Attention Generative Adversarial Networks for Cross-domain Senti. Class.~\cite{Zhang2019} 
& Conditional BERT Contextual Augmentation~\cite{Wu2019} 
& Inductive Hashing on Manifolds~\cite{Shen2013}
& Learning Representations for Senti. Class. using Multi-task framework~\cite{Meisheri2018} \\

\hline
\end{tabular}
}

\end{table*}

\textit{\citeauthor{Gupta2019} \cite{Gupta2019}} is the seed paper to which \pwc~associates three \textit{task} labels (\textit{data augmentation}, \textit{sentiment analysis}, \textit{text generation}), two \textit{method} labels (\textit{convolution} and \textit{generative models (GAN)}), and none \textit{dataset} label.
As the labels and the title suggests, \textit{\citeauthor{Gupta2019} \cite{Gupta2019}} uses generative adversarial networks as a data augmentation method to generate textual training data for the sentiment classification task.
The four different recommendation sets illustrate the many facets in that papers can be related.

The generic recommendations are all about GAN as an augmentation method. 
While the first and third recommendations \textit{\citeauthor{Karimi2020} \cite{Karimi2020}} and \textit{\citeauthor{Zhang2019} \cite{Zhang2019}} are both also about sentiment classification, the second \textit{\citeauthor{Zhu2018} \cite{Zhu2018}} investigates emotion classification. 
Even though sentiment and emotion can be considered as related, the former is based on text and the latter on image data.

All \textit{task}-specific recommendations \textit{\citeauthor{Anaby-Tavor2019} \cite{Anaby-Tavor2019}}, \textit{\citeauthor{Regina2020} \cite{Regina2020}}, and \textit{\citeauthor{Wu2019} \cite{Wu2019}} have data augmentation on text classification as a central theme.
However, in contrast to the seed, GANs are not used for augmentation, and the classification task is not concerned with sentiment. %
The \textit{method}-specific recommendations \textit{\citeauthor{Zahan2020} \cite{Zahan2020}}, \textit{\citeauthor{Husmann2020} \cite{Husmann2020}}, and \textit{\citeauthor{Shen2013} \cite{Shen2013}} are at first sight unrelated to the seed since they focus on unrelated topics such as hashing or the classification of biomedical or financial data.
Nonetheless, the seed and the \textit{method}-specific recommendation all use t-distributed Stochastic Neighbor Embedding (t-SNE) for visualization.
Despite of being different in central themes, the paper pairs have similar methodologies. %
The similarity between the seed and the \textit{dataset}-specific recommendations is evident.
\textit{\citeauthor{Gupta2018} \cite{Gupta2018}}, \textit{\citeauthor{Xiang2020} \cite{Xiang2020}}, and \textit{\citeauthor{Meisheri2018} \cite{Meisheri2018}} are all about sentiment classification in low resource settings.
Instead of data augmentation with GAN, they utilize external knowledge or transfer learning.

In summary, we consider all recommendations as generally relevant since they share one or more aspects with the seed.
Due to the subjectiveness of relevance, a recommender system would need to relate the recommendations to its users' individual information needs.
However, when new user data is unavailable, this is not feasible. %
This is a general problem of purely content-based recommendations.
Our sample example illustrates how different aspects can approximate similar research papers in a granular and more detailed perspective.
The specialization from Siamese-SciBERT also leads to diverse recommendations between aspect-specific recommendations and generic ones.
SPECTER's generic recommendations have a relatively narrow focus on data augmentation with GAN for classification.
The \textit{method}-specific recommendations even reveal the implicit shared use of the t-SNE visualization.

\section{Discussion}

Our quantitative and qualitative results reveal the effect of specialized document embeddings.
The performance gains between the best generic and the best specialized embeddings,
i.\,e., generic SPECTER and Siamese-SciBERT, are substantial.
We anticipated this outcome as the generic embeddings are not optimized for this task compared to the specialized ones.
Still, our findings do not mean generic embeddings lead to unrelated recommendations, but only that they are not similar concerning \textit{task}, \textit{aspects}, or \textit{dataset}.
Siamese-SciBERT also outperforms the Pairwise SPECTER baseline.
The pairwise SPECTER with a unbounded $n$ may yield better results than our baseline implementation. 
However, due to the quadratic complexity, we have to perform ~1.3 billion comparisons, which would take approximately 46 days on the hardware used in our experiments (GeForce RTX 2080 Ti with 11GB memory).
Thus, the potential performance gains do not justify the increase in computational effort.

\paragraph{Specialization Performance}
In terms of specialization, the Siamese Transformer (Siamese-SciBERT) outperforms retrofitting and non-Siamese Transformer fine-tuning. %
This outcome can be explained by several reasons.
The retrofitting method from \citet{Glavas2018} has been originally developed for words and optimized for the properties of a word embedding space.
We see retrofitting has a positive effect on Avg.~FastText but a negative effect on SPECTER.
SPECTER uses citation information and, therefore, its embedding space has different properties~\cite{Cohan2020}.
At the same time, SPECTER's citation information generally improves the performance of its generic and fine-tuned version compared to SciBERT. 
The poor performance of SciBERT is aligned with the results of related studies \cite{Reimers2019,Ostendorff2020}, i.\,e., document embeddings from BERT-based models are unsuited for the similarity search.
Since we perform the similarity search based on static embeddings, each document needs to be independently encoded.
While this is the case in Siamese-SciBERT, the non-Siamese Transformers (SPECTER and SciBERT) are fine-tuned in the sequence pair classification setting, i.\,e., a document pair is jointly encoded. 
As the results from \cite{Ostendorff2020} suggest, the joint encoding is superior for pairwise document classification approach.
However, our results show the opposite in a similarity search setting.
The independent encoding, as in the Siamese model, produces semantically similar documents embeddings with higher precision and recall. 

Given the overall results, we consider Siamese-SciBERT as the best tested method to specialize embeddings. 
Nevertheless, we ask ourselves if the specialization effect depends on individual aspects.
The most positive specialization effect can be observed for the \textit{method} aspect, while the effect is less significant for \textit{dataset}.
We partially attribute the discrepancy in the specialization effect to training data availability, e.\,g., more samples for \textit{method} than \textit{dataset}.
However, the effect is also due to the aspects being differently inherent in generic embeddings' similarity.

\paragraph{Bias in Generic Embeddings}

The similarity of generic embeddings does not explicitly contain aspect information, i.\,e., we cannot attribute the document similarity to a specific aspect in which documents are similar.
However, we can assume the aspects are implicitly part of the similarity.
Thus, the similarity of generic embeddings would be denoted as a weighted sum 
$\sum_{a\in A}w_a*s_a $, where $A=\{\textrm{\textit{task}}, \textrm{\textit{method}}, \textrm{\textit{dataset}}, \dots a_n\}$ is a set of aspects consisting of our three and an arbitrary number of other aspects.
If the similarity of generic embeddings would evenly incorporate all aspects, all weights $w_a$ should be equal.
Still, our experiments suggest the aspects are not equally weighted.
Table~\ref{tab:overlap} reports an uneven intersection ratio among the recommendations.
The \textit{method}-specific recommendations have less overlap with the generic recommendation than the \textit{dataset} or \textit{task}-specific recommendations.
Given that \textit{task} has the most samples in the ground truth, we would have expected a different outcome, e.\,g., more specialization concerning \textit{task}.
Therefore, $w_{\textrm{\small\textit{method}}} < w_{\textrm{\small\textit{task}}} < w_{\textrm{\small\textit{dataset}}}$ likely holds true.
Accordingly, the results indicate an implicit bias in the similarity of generic embeddings towards \textit{dataset} and against \textit{method}.
Our qualitative analysis does not reject this finding.
We hypothesize the bias is more likely to be caused by the corpus' characteristics than by the embedding methods themselves.
Title and abstract of papers prominently mention tasks and datasets, whereas methodological details are of marginal importance, e.\,g., the t-SNE visualization in our example from Table~\ref{tab:examples}.

\paragraph{Implications for Content-based Recommender Systems}

Having this bias towards a single aspect indicates the generic embeddings present only a single view on the content of a document.
Therefore, the conflation of meaning, which have been shown for word embeddings~\cite{Pilehvar2016,Camacho-Collados2018}, also exists for document embeddings.
Consequently, a recommender system based on the generic embeddings is limited in the information needs that the system can address.
Namely, those information needs that match with the single aspect, which is in our case the \textit{dataset} aspect. %
Such a narrow focus on one information need hurts the diversity of the recommendations.
In the literature~\cite{Nguyen2014,Ge2010}, the lack of diversity has been identified as a major issue of today's recommender systems.
By changing the approach of representing documents, from generic to specialized embeddings, diverse information needs can be addressed even when user data is sparse.
In the context of recommendations, our data does not allow a decisive  statement on the relevancy of the generic or aspect-based recommendations since we primarily evaluate the similarity of research papers. %
We use similarity only as an approximation of relevance for specific information needs, i.\,e., interest in the task, method, or dataset of the presented research. 
To the best of our knowledge, a dataset that would allow a relevance-based evaluation of the \pwc corpus is not publicly available.
Thus, further experiments involving user feedback are required to investigate the relevancy of aspect-based recommendations.
Nonetheless, the recommendations from specialized embeddings can expose the implicit bias within the generic recommendations.
Integrating the aspect information can improve research paper recommender systems as users would decide in which particular aspect they are interested.
Thereby, tailored content-based recommendations are feasible even without user feedback.
The aspect-based recommendation would increase the transparency of a recommender system since the system could provide explicit explanations on the aspects in that documents are related.
Such explanations would also strengthen the trust in the recommendations as \citeauthor{Kunkel2019}~\cite{Kunkel2019} demonstrate.
Furthermore, diversity can be addressed through selection from multiple aspects.
In a user interface, one would not only display recommendations from a particular aspect but rather select one recommendation from each aspect, e.\,g., the top recommendation for \textit{task}, \textit{method}, and \textit{dataset} (the items in the first row of Table~\ref{tab:examples}). 

\paragraph{Scalability}

Diversity and explainability are also covered by the pairwise multi-class classification approach \citeauthor{Ostendorff2020c} \cite{Ostendorff2020c}.
However, the pairwise approach bears scalability constrains that would prevent recommender systems to be deployed in a production environment. 
Pairwise document classification requires large computational resources even for medium-sized corpora since aspect information need to be separately derived for all document pairs. 
To use the pairwise approach as a baseline, we introduced the candidate filtering but it still needs to perform $11.3$M Transformer forward-passes while achieving only a lower performance compared to Siamese-SciBERT.
Instead, our approach derives the aspect information during the encoding phase, which results in a linear complexity  (118,146 forward-passes in our experiments).
During the indexing of a new document, the system would only need to create $n$ specialized embeddings instead of a single generic embedding.
Thus, our approach's complexity is mainly bound to the number of aspects and not to the size of the document corpus as in pairwise classification (see Section~\ref{ssec:approach}).
As a result, our approach is applicable for real-world recommender systems on commodity hardware. 
Our Web-based demo is one example for prototypical recommender system based on specialized document embeddings\footref{fn:demo}.

\paragraph{Interpretability}

Aside from scalability, the specialized embeddings have additional advantages such as explainablility and interpretability. 
Each individual aspect-specific vector $\vec{d}_i^{(a_j)}$ could also be combined through concatenation into a single document vector $\vec{d}_i=[\vec{d}_i^{(a_1)};\dots;\vec{d}_i^{(a_n)}]$ for other downstream tasks.
The aspect's dimensions could then facilitate the interpretability of the document vectors in similar fashion as \citet{Liao2020} already demonstrated with sparse vectors.
In the context of words, related approaches already exist.
For example, \citet{Schwarzenberg2019} project word vectors into a concept space in which the dimensions correspond to predefined concepts. %

\paragraph{Alternative Approaches}

Lastly, the question is whether comparable recommendations are also possible with alternative approaches such as query-sensitive similarity~\cite{Tombros2001}.
One could filter papers by a query, i.\,e., their respective aspect labels, and then perform a nearest neighbor search on the filtered papers' generic embeddings.
However, the filtering depends on hard label assignments, e.\,g., papers need to have an identical task, method, or dataset to be considered.
Papers only similar in a particular aspect would be excluded.
In our example (Table~\ref{tab:examples}), \textit{\citeauthor{Zhu2018}~\cite{Zhu2018}} would have been excluded because its task is \textit{emotion classification} related but not identical with \textit{sentiment classification} as in the seed document.
Moreover, the specialized embedding space allows dissimilarity search, e.\,g., considering papers with similarity above a certain threshold.
This allows retrieving papers similar in their task but different in their method.
The formulation of such queries could furthermore facilitate the discovery of analogies between research papers \cite{Chan2018}.

\section{Conclusions}
This paper introduces our approach of specialized document embeddings for aspect-based document similarity of research papers.
Instead of considering each research paper as a single entity for document similarity, we incorporate multiple aspects in our approach, i.\,e., \textit{task}, \textit{method}, and \textit{dataset}.
Therefore, we move from a single generic representation to three specialized ones.
We treat aspect-based similarity as a classical vector similarity problem in aspect-specific embedding spaces.
Our approach contributes two major improvements over existing literature of aspect-based document similarity: %
In contrast to segment-level similarity \cite{Chan2018,Kobayashi2018,Huang2020}, a document is not divided into segments which harms the coherence of a document.
Instead, we preserve the semantics of the whole document that are needed for a meaningful representation. 
Additionally, our approach is less resource intensive and achieves a higher precision and recall compared to the pairwise document classification baseline \cite{Ostendorff2020,Ostendorff2020c}.
The improved scalability allows the development a real-world recommender system, which we demonstrate with our demo\footref{fn:demo}. 

In our empirical study, we compare and analyze three generic document embeddings, six specialized document embeddings and a pairwise classification baseline in the context of research paper recommendations. 
To the best of our knowledge, all applied specialization methods were, so far, used only to derive generic embeddings.
Our evaluation is conducted on the newly constructed \pwc corpus containing more than $150,000$ research papers.
This \pwc corpus is unique for research on aspect-based document similarity as it contains manual annotations regarding different aspects of research papers. 
In our experiments, Siamese-SciBERT %
outperforms all other methods with 0.224 MAP for \textit{task}-, 0.137 MAP for \textit{method}-, and 0.235 MAP for \textit{dataset}-specific recommendations.
Our comparison between recommendations using generic and specialized embeddings indicates a tendency of generic recommendations being more similar regarding \textit{dataset} than \textit{method}.
Thus, papers with a similar method are less likely to be recommended with these generic embeddings.
Our approach of aspect-based document embeddings mitigates potential risks arising from implicit biases by making them explicit. 
This can, for example, be used for more diverse and explainable recommendations, e.\,g., by recommending documents for every aspect.
The development of an aspect-based recommender system and its evaluation with user feedback is subject to future work.

\bibliographystyle{ACM-Reference-Format}
\bibliography{anthology,custom,example_recommendations}

\end{document}